\documentclass[english,a4paper]{article}
\usepackage{dcolumn}
\usepackage{bm}
\usepackage{graphicx}
\usepackage{amsmath}
\usepackage{amsfonts}
\usepackage{amssymb}
\usepackage{amsthm}
\usepackage{amstext}
\usepackage{amsbsy}
\usepackage{amsopn}
\usepackage{amscd}
\usepackage{amsxtra}
\usepackage{mathtools}
\usepackage{commath}

\def\openone{\leavevmode\hbox{\small1\kern-3.3pt\normalsize1}}

\usepackage[usenames,dvipsnames]{color}

\def\phi{\varphi}
\def\epsilon{\varepsilon}

\begin{document}

\title{Robust control of a NOT gate by composite pulses}

\author{G. Dridi\footnote{Laboratoire Mat\'eriaux Avanc\'es et Ph\'enom\`enes Quantiques, D\'epartement de physique, Facult\'e des Sciences de Tunis, Universit\'e de Tunis El Manar, Compus universitaire, Tunis 1068}, M. Mejatty\footnote{Institut Sup\'erieur des Sciences Appliqu\'ees et de Technologies de Gafsa, Université de Gafsa, Campus universitaire Sidi Ahmed Zarroug, 2112, Gafsa}, S. J. Glaser\footnote{Department of Chemistry, Technische Universit\"at M\"unchen, Lichtenbergstrasse 4, D-85747 Garching, Germany}, D. Sugny\footnote{Laboratoire Interdisciplinaire Carnot de
Bourgogne (ICB), UMR 6303 CNRS-Universit\'e Bourgogne-Franche Comt\'e, 9 Av. A.
Savary, BP 47 870, F-21078 Dijon Cedex, France, dominique.sugny@u-bourgogne.fr}}

\maketitle

\begin{abstract}
We present a general procedure to implement a NOT gate by composite pulses robust against both offset uncertainties and control field variations. We define different degrees of robustness in this two-parameter space, namely along one, two or all directions. We show that the phases of the composite pulse satisfy a nonlinear system, and can be computed analytically or numerically.
\end{abstract}

\section{Introduction}\label{sec1}
Quantum control ~\cite{Allen:87, Brif:10, B.W.Shore, M.Shapiro, Nielsen} is nowadays a promising technique with a wide range of applications in a multitude of domains extending from molecular physics~\cite{M.Shapiro,RMP:rotation}, Nuclear Magnetic Resonance (NMR)~\cite{ota:2009, Levitt:1982, E.L.Hahn, M.H.Levitt, R.Freeman,lapert2012} and more recently quantum information science~\cite{Nielsen, vandamme:2017, garon:2013, ivanov:2015, torosov:2014, torosov:2011b, cohen:2016, ichikawa:2014, wang:2014, jones:2013, li:2006, morton:2005, hugh:2005, jones:2003, low:2016}. Quantum control is aimed at bringing the state of the system towards a given target state by means of external electromagnetic fields. In quantum computing, this transfer concerns the propagator of the dynamics and has to be realized with a very high efficiency and in a robust manner with respect to experimental uncertainties and variations of the control field. Different methods have been explored up to date to design the corresponding electric or magnetic excitations. Optimal control techniques~\cite{Glaser:15,wang2012,owrutsky2012,skinner2003,Nimbalkar2013,kobzar2005,turinici2004,kobzar2004,vandamme:2017b,assemat:2010,Ghassen1,reich2012,bonnard2012} are actively developed for this purpose, but generally only numerical solutions are designed. Adiabatic~\cite{RMPvitanov,S.Guerin1,N.V.Vitanov1,baum:1985,Ghassen2} and shortcut to adiabaticity  pulses~\cite{STAreview,Xi,daems:2013,ruschhaupt2012} have been also the subject of an intense activity in this direction. Another historical solution to this problem, already proposed in the early eighties in NMR, is the use of composite pulses (CP), i.e. a sequence of resonant pulses with specific phases computed to compensate the errors of the control process~\cite{jones:2013,ota:2009, Levitt:1982, E.L.Hahn, M.H.Levitt, R.Freeman, levitt:1986, levitt:1979, levitt:1981,levitt:1986}. CP or some extensions such as the Schinnar Le Roux algorithm~\cite{SLR91} are by now widely used in NMR, medical imaging and quantum computing. The interest on CP has been recently renewed by a series of papers showing the generality, the versatility and the universality of the CP approach~(see Ref.~\cite{BM:2014,jones:2013,genov:2014,torosov:2018,dou:2016,ivanov:2015,ota:2009,low:2014,torosov:2015,torosov:2014,kyoseva:2013, torosov:2013, vitanov:2011, torosov:2011,torosov:2011b,cohen:2016,merrill:2017,kabytayev:2014,brown:2004} to mention a few). In this setting, a large amount of studies focuses on simple control tasks~\cite{genov:2014}, such as robust population inversion. Other investigations have performed more complex goals, e.g. the implementation of robust qubit gates~\cite{ichikawa:2011,jones:2013,wang:2014}. A specific target state, namely the NOT gate, is generally used. CP have been extended to the case where off-resonant detunings are used as control parameters~\cite{CPdetuning}. Some studies consider only one source of imperfections either on the system (offset terms)~\cite{SLR91} or on the external field~\cite{low:2016}. Different techniques have been introduced to compute the parameters of CPs.

We propose in this paper to revisit the previous studies on the subject. We introduce a general framework to derive robust CPs for the implementation of quantum gates. We consider as an illustrative example the NOT gate. Two standard sources of imperfection are accounted for, i.e. the offset uncertainty and the control field inhomogeneity. They are each described by an inhomogeneous and constant parameter in the dynamical equation~\cite{kobzar2004,daems:2013}. We introduce different degrees of robustness, namely along a line, two lines or along all directions in this two-parameter space. In each case, we show, without any approximation, that the phases of the CP satisfy a nonlinear system. This system can be solved either analytically for the simplest CPs or numerically. We discuss the limitations of this approach in terms of complexity and efficiency. In particular, we show that a symmetric CP is sufficient to be robust along a line, while, for two lines, a general CP has to be used. We also propose a procedure to be robust along all directions, and we analyze the extent to which this robustness can be performed. A comparison with other CPs derived in the literature is finally made.

The remainder of the paper is organized as follows. In Sec.~\ref{sec2}, we describe the model system used to  design CPs  robust against both offset uncertainties and control field variations. In Sec.~\ref{sec3},  we outline the principles of the general method, paying special attention to its flexibility and  applicability. The NOT gate is taken as an illustrative example. Conclusion and prospective views are given in Sec.~\ref{conc}. Technical details are reported in Appendix~\ref{coeffapp}.

\section{The model system}\label{sec2}
We consider a two-level quantum system driven by a composite sequence of $N$ identical pulses, with different phases $\phi_n$. The CP has a total duration $NT$, where $N$ is the number of phases and $T$ the duration of each individual pulse. The pulse of amplitude $\omega_0$ is given between the times $nT$ and $(n+1)T$ by:
\begin{equation}
\omega_{x,n}=\omega_0 \cos (\phi_n), \quad \omega_{y,n}=\omega_0 \sin (\phi_n).
\end{equation}
In presence of field inhomogeneities, the dynamics of the system are governed by the Hamiltonian:
\begin{equation}
H_n=-\frac{1}{2}\left[ \begin{matrix}
 \delta &(1+\eta)(\omega_{x,n}-i\omega_{y,n})\\
 (1+\eta)(\omega_{x,n}+i\omega_{y,n}) &-\delta
  \end{matrix}\right],
 \end{equation}
where units such that $\hbar=1$ are used throughout the paper. The modelling of the imperfect knowledge of the quantum system is described by the two parameters $\delta$ and $\eta$, which correspond respectively to an unknown offset (due to an
inhomogenous broadening or to an imperfect driving frequency) and to a scaling factor associated with variations in the field amplitude. The phases $\phi_n$ are designed to build a pulse sequence robust against these two inhomogeneous parameters. Introducing the normalized vector $\vec{\omega}_n$:
\begin{equation}
 \vec{\omega}_{n}=\frac{1}{\omega}\left(\begin{array}{ccc}
 (1+\eta)\omega_{x,n}\\
(1+\eta)\omega_{y,n} \\
 \delta
\end{array}  \right),
\end{equation}
where $\omega=\sqrt{\delta^{2}+\omega_0^2(1+\eta)^{2}}$, the Hamiltonian $H_n$ can be written as:
\begin{equation}
H_n=-\frac{\omega}{2}\vec{\omega}_{n}\cdot \vec{\sigma},
\end{equation}
where the components of the vector $\vec{\sigma}$ are the Pauli matrices. The propagator $U_n$ of the system between the times $nT$ and $(n+1)T$ is then given by:
\begin{equation}\label{eqpropa}
U_n=e^{-iH_n T}=\cos\left( \frac{\omega T}{2}\right)\mathbf{1}+i\sin\left( \frac{\omega T}{2}\right) \vec{\omega}_{n}\cdot \vec{\sigma},
\end{equation}
with $\mathbf{1}$ the Identity matrix. Using the variables:
\begin{equation}
\begin{cases}
x = \frac{(1+\eta)\omega_0}{\omega}\sin\left( \frac{\omega T}{2}\right)\\
y = \frac{\delta}{\omega}\sin\left( \frac{\omega T}{2}\right)\\
z = \cos\left( \frac{\omega T}{2}\right),
\end{cases}
\end{equation}
the propagator \eqref{eqpropa} reads:
\begin{equation}\label{eq_pq}
U_n=\left( \begin{matrix}
 z+iy&ixe^{-i\phi_{n}}\\
 ixe^{i\phi_{n}} &z-iy
\end{matrix}\right)=\left( \begin{matrix}
  q e^{i\alpha}&ipe^{-i\phi_{n}}\\
 ipe^{i\phi_{n}} &q e^{-i\alpha}
\end{matrix}\right),
\end{equation}
with $q=\sqrt{y^{2}+z^{2}}$, $p=x$, and $\alpha=\arctan (y/z)$~\cite{genov:2014}. Note that the parameters $\alpha$, $q$ and $p$ only depend on $\delta$ and $\eta$ and that they satisfy the relation $p^{2}+q^{2}=1$. A sequence of $N$ pulses with the same pulse amplitude but with different phases $\phi_{n}$ produces the propagator $U^{(N)}$ which can be expressed as:
$$
U^{(N)}=U_NU_{N-1}\cdots U_1.
$$

A general target evolution operator in SU(2) can be expressed as:
\begin{equation}
U_{t}=\left( \begin{matrix}
  a&-b^{ \ast}\\
  b&a^{ \ast}
 \end{matrix}\right),
\end{equation}
where $a$ and $b$ are two complex numbers such that $\vert a \vert^{2}+\vert b\vert^{2}=1$. The control objective is to maximize the figure of merit $J(\delta,\eta)$ in a neighborhood of $\delta=\eta=0$:
 \begin{eqnarray}\label{figm}
 \nonumber J(\delta,\eta)&=& \frac{1}{2} \text{Tr}\left(U_{t}^{\dagger} U^{(N)} \right)\\
\nonumber &=&\Re(a)\Re(U^{(N)}_{1,1})+\Im(a)\Im(U^{(N)}_{1,1})\\
 &+&\Re(b)\Re(U^{(N)}_{2,1})+\Im(b)\Im(U^{(N)}_{2,1}),
\end{eqnarray}
where $U^{(N)}_{1,1}$ and $U^{(N)}_{2,1}$ are two matrix elements of $U^{(N)}$. Using Eq.~\eqref{eq_pq}, we obtain:
\begin{equation}
\begin{cases}
\Re{\left( U^{(N)}_{1,1}\right)}=\sum_{k=0}^{\frac{N-1}{2}} A_{k}p^{2k}q^{N-2k}\\
\Im{\left( U^{(N)}_{1,1}\right)}=\sum_{k=0}^{\frac{N-1}{2}} B_{k}p^{2k}q^{N-2k}\\
\Re{\left( U^{(N)}_{2,1}\right)}=\sum_{k=0}^{\frac{N-1}{2}} C_{k}q^{2k}p^{N-2k}\\
\Im{\left( U^{(N)}_{2,1}\right)}=\sum_{k=0}^{\frac{N-1}{2}} D_{k}q^{2k}p^{N-2k}
\end{cases}
\end{equation}
where the coefficients $A_{k}$, $B_k$, $C_k$ and $D_k$ depend on $\alpha$ and the angles $\phi_n$. The general expression of the coefficients is given in Appendix~\ref{coeffapp} for $N=5$. We observe that the figure of merit can be expressed as a polynomial of degree $N$ in $p$ and $q$. In this paper, we investigate the case of the NOT gate whose propagator is given by:
  \begin{equation}\label{Eq.11}
 U_{t}=\left( \begin{matrix}
  0 & 1\\
  -1 & 0
  \end{matrix}\right).
  \end{equation}
This leads to $\Re(a)=\Im(a)=\Im(b)=0$ and $\Re(b)=-1$. The figure of merit can now be expressed as:
\begin{eqnarray}\label{eq12}
\nonumber J(\alpha, p, q)=-\Re{\left( U^{(N)}_{2,1}\right)}&=&\sum_{k=0}^{\frac{N-1}{2}} c_{k}q^{2k}p^{N-2k},
\end{eqnarray}
with $c_k=-C_k$. In Sec.~\ref{sec3}, we consider an odd number of composite $\pi$- pulses such that $\omega_{0}=\pi/T$. In the absence of inhomogeneities (i.e. $\delta=\eta=0$), this pulse sequence gives $J=1$ and leads to a perfect NOT gate~\cite{torosov:2015, ivanov:2015, torosov:2014}.
\section{The general approach}\label{sec3}
\begin{figure}[!h]
\includegraphics[scale=0.4]{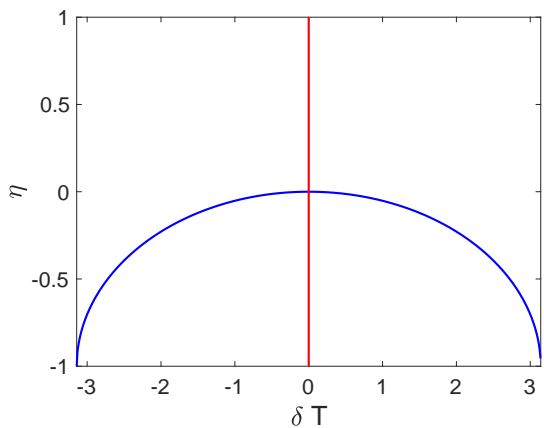}
   \caption{ (Color online)  Plot of the lines of equations $\delta=0$ (red or gray) and $\eta=\sqrt{1-\delta^2T^2/\pi^2}-1$ (blue or dark gray) in the space $(\delta T,\eta)$. Dimensionless units are used.}
  \label{fig.1}
  \end{figure}

This paragraph is aimed at exploring how to design the phases $\phi_n$ in order to implement a robust NOT gate against variations in $\delta$ and $\alpha$. We show that different types of robustness can be considered in this two-parameter space. We first point out that the coefficients of the polynomial $J$ cannot be chosen freely since they depend on the inhomogeneous parameters $(\delta,\eta)$ through $\alpha$. In the space $(\delta,\eta)$, we introduce two lines of equations $\delta=0$ and $\delta^2T^2+\pi^2(1+\eta^2)=\pi^2$, which are displayed in Fig.~\ref{fig.1}. Since $\alpha=\arctan (y/z)$, the two lines are associated respectively to the values $\alpha=0$ and $\alpha=\pi/2$. Note that the two lines intersect in the point $\delta=\eta=0$. Other lines characterized by a constant value of $\alpha$ could be also considered. In a neighborhood of $\delta=\eta=0$, straightforward computations lead to $\tan\alpha\simeq \frac{2T}{\pi^2}\frac{\delta}{\eta}$, which shows that any direction in the $(\delta,\eta)$- space can be chosen.

\begin{figure}[!h]
\includegraphics[scale=0.4]{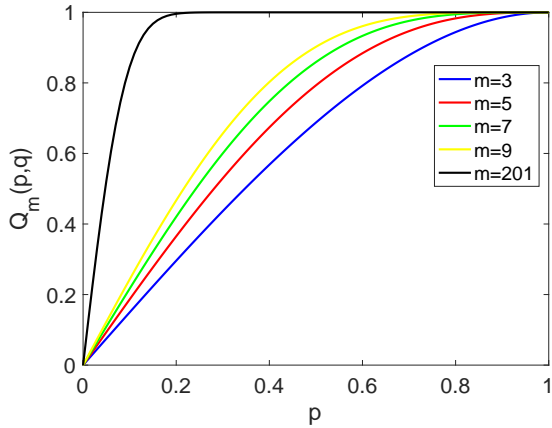}
   \caption{ (Color online) Polynomial $Q_m$ defined in Eq.~\eqref{eqnewton} as a function of the parameter $p$ for $m=3$ (blue or dark gray), 5 (red or gray), 7 (green or light gray), 9 (yellow or white) and $m=201$ (black). Dimensionless units are used.}
  \label{fig.2}
  \end{figure}

We propose a general procedure to design robust pulses along one or both of these lines. The point $\delta=\eta=0$ corresponds to $p_0=p(\delta=0,\eta=0)=\sin(\frac{\omega_0T}{2})=1$ and to $q_0=q(\delta=0,\eta=0)=\cos(\frac{\omega_0T}{2})=0$. A relevant figure of merit $J$ requires that:
\begin{equation}\label{eqpoly}
|J(\alpha,p,q)|\leq 1
\end{equation}
in a neighborhood of $(p_0,q_0)$. A systematic solution is given by the following polynomial $Q_m$ of order $m$ in $p$:
\begin{equation}\label{eqnewton}
Q_m(p,q)=\sum_{k=0}^{\frac{m-1}{2}}c_kq^{2k}p^{m-2k},
\end{equation}
where
\begin{equation}\label{coeff}
c_k=\left( \begin{matrix}
  m/2\\
  k
  \end{matrix}\right)=\frac{m/2 (m/2-1) \cdots (m/2-k+1)}{k!}.
\end{equation}
Different polynomials $Q_m$ for $m=3$, 5, 7, 9 and $m=201$ are plotted in Fig.~\ref{fig.2}. It can be shown that the $(N-1)/2$ first derivatives of $Q_m$ with respect to $p$ at $p=p_0$ are zero. Note that $Q_m$ has the advantage to satisfy the constraint \eqref{eqpoly} for $p\in [0,1]$. The approach proposed in this study can be applied to other polynomials which fulfill Eq.~\eqref{eqpoly} in a neighborhood of $(p_0,q_0)$.

\subsection{Robustness along the line $\alpha=0$ }\label{sec3a}
\begin{figure}[!h]
\includegraphics[scale=0.3]{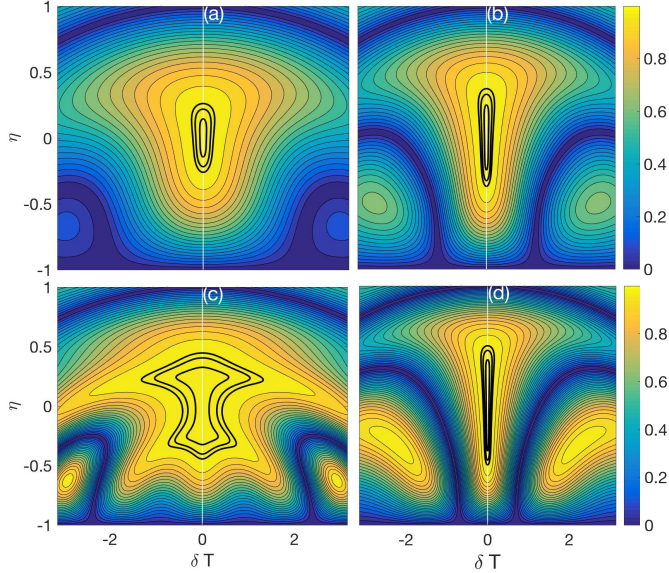}
   \caption{(Color online) Fidelity $\vert J(\delta,\eta) \vert$ as a function of the offset $\delta$ and the scaling factor $\eta$ for the case $\alpha=0$. Panels (a), (b), (c) and (d) correspond respectively to a composite sequence with 3, 5, 7 and 9 pulses. The black bold solid lines (from inside to outside) correspond respectively to the level lines $\vert J(\delta,\eta) \vert$=0.999, 0.995 and 0.99. The equation of the white solid line is $\delta=0$. Dimensionless units are used.}
        \label{fig3}
  \end{figure}

We consider in this paragraph the case $\alpha=0$. The polynomial $J$ has $(N+1)/2$ coefficients, with $N$ phases $\phi_n$ to determine. The number of constraints being smaller than the number of phases, a simple solution can be derived by using a symmetric CP in which the phases satisfy the anagram relation $\phi_{N+1-k}=\phi_{k}$, with $k=1,2,\cdots,(N-1)/2$ ~\cite{torosov:2015, ivanov:2015, torosov:2014}.  We assume here that the polynomial $J(p,q)$ is given by $Q_m$ with $N=m$. As a first illustrative example, we consider the case $N=3$.  Using Eq.~\eqref{coeff}, we arrive at:
\begin{equation}\label{N3}
J_(p,q,\alpha)=c_0p^3+c_1q^2p,
\end{equation}
with $c_0=1$, $c_1=3/2$ and
\begin{equation}\label{eqjp}
\begin{cases}
c_0=\sin(-\phi_1-\phi_3+\phi_2) \\
c_1= \sin(2\alpha+\phi_3)+\sin(\phi_2)-\sin(2\alpha-\phi_1).\\
\end{cases}
\end{equation}
The phases are then computed by solving this nonlinear system for $\alpha=0$ and $\phi_1=\phi_3$. An exact solution is  given by: $\phi_1=\phi_3=\pi/6$ and $\phi_2=5\pi/6$. Following this approach, we have determined the phases for $N=5$, $7$ and $9$. The solutions are given in Tab.~\ref{Table1}. A numerical solver has been used to find the phases with a very good accuracy for $N\geq 5$. The efficiency of the corresponding control strategy is shown in Fig.~\ref{fig3} for $N=3$, 5, 7 and 9. As could be expected, we observe that the robustness is maximum in a narrow region around the line $\delta=0$. Note the surprising result obtained for $N=7$, with a robustness covering a wider area.
\begin{table}[htp]
\caption{Values of the phases which maximize the robustness along the line $\alpha=0$ for the case $N=3$, 5, 7 and 9.}
\begin{center}
\begin{tabular}{|l|c|}
 \hline
  \hline
  $N $& $\phi_k, k=1...N $\quad (rad)\\
\hline
  3 & $\pi/6,  5\pi/6, \pi/6$  \\
 \hline
      5 &$0.211, 1.21, 3.569, 1.21, 0.211$  \\
    \hline
       7 &$0.706, -0.903, 1.98, 2.466, 1.98, -0.903, 0.706 $ \\
 \hline
   9&$0.03, 0.242, 0.941, 2.466, 5.043, 2.466, 0.941, 0.242, 0.03$ \\
     \hline
      \hline
       \end{tabular}
\end{center}
\label{Table1}
\end{table}
\subsection{Robustness along the line $\alpha=\pi/2$ }\label{sec3b}
\begin{figure}[!h]
\includegraphics[scale=0.3]{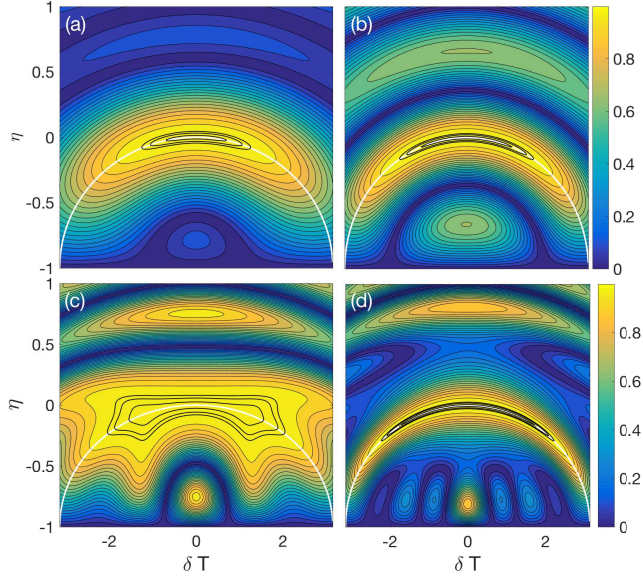}
   \caption{(Color online) Same as Fig.~\ref{fig3} but for $\alpha=\pi/2$. The equation of the white solid line is $\eta=\sqrt{1-\delta^2T^2/\pi^2}-1$.}
  \label{fig4}
  \end{figure}

The same technique can be used when the parameter $\alpha$ is set to $\pi/2$. The figure of merit $J$ is still given by the polynomial $Q_m$. For the case $N=3$, the phases are solutions of the system~\eqref{eqjp} with $\alpha=\pi/2$. In the symmetric configuration $\phi_1=\phi_3$, an exact solution is $\phi_1=\phi_3=\pi/6$ and $\phi_2=-\pi/6$. The same analysis can be done for $N=5$, $7$ and $9$. Exact solutions have been found for $N=5$. Very good numerical estimates are given for $N=7$ and 9. The corresponding values are presented in Tab.~\ref{Table2}. Figure~\ref{fig4} displays the robustness along the line $\eta=\sqrt{1-\delta^2T^2/\pi^2}-1$ of the different pulse sequences. Here again, we observe a remarkable robustness for $N=7$.
\begin{table}[htp]
\caption{Same as Tab.~\ref{Table1} but for $\alpha=\pi/2$.}
\begin{center}
\begin{tabular}{|l|c|}
 \hline
 \hline
   $N $& $\phi_k, k=1...N $ \quad (rad) \\
  \hline
  3 & $\pi/6,  -\pi/6, \pi/6$  \\
   \hline
     5 &$\pi/14, -3\pi/5, -59\pi/70,-3\pi/5, \pi/14$  \\
     \hline
       7 &$-0.691, -2.257, -2.002, -5.584, -2.002, -2.257, -0.691 $ \\
    \hline
  9&$1.012, 0.442, 2.569, 2.945, 1.184, 2.945, 2.569, 0.442, 1.012$ \\
     \hline
     \hline
      \end{tabular}
\end{center}
\label{Table2}
\end{table}

\subsection{Robustness along both lines $\alpha=0$ and $\alpha=\pi/2$ }\label{sec3c}
\begin{figure}[!h]
  \includegraphics[scale=0.3]{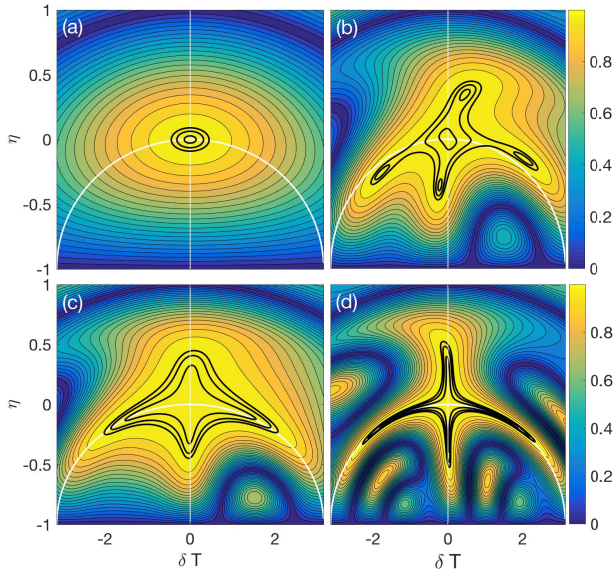}
   \caption{(Color online) Same as Fig.~\ref{fig3} but for both lines $\alpha=0$ and $\alpha=\pi/2$. The equation of the two white lines are $\delta=0$ and $\eta=\sqrt{1-\delta^2T^2/\pi^2}-1$.}
  \label{fig5}
  \end{figure}

In this paragraph, we design a pulse sequence robust along both lines $\alpha=0$ and $\alpha=\pi/2$. The figure of merit is still given by Eq.~\eqref{eqnewton}. Each coefficient of the polynomial must verify: $c_k(\alpha=0)=c_k(\alpha=\pi/2)$, $k=1,\cdots,(N-1)/2$.  This generates a system of $N$ equations, with $N$ phases to determine. Here, due to the additional constraints, we cannot consider a symmetric CP. In the case $N=3$, we get:
  \begin{equation}\label{eqjp2}
\begin{cases}
c_0=\sin(-\phi_1-\phi_3+\phi_2)=1 \\
c_1(\alpha=0)= \sin(\phi_3)+\sin(\phi_2)+\sin(\phi_1)=3/2\\
c_1(\alpha=\pi/2)= -\sin(\phi_3)+\sin(\phi_2)-\sin(\phi_1)=3/2\\
\end{cases}
\end{equation}
In general, this nonlinear system has not an exact solution and only approximate solutions can be found. For $N = 3$, the best choice of phases is $\phi_1=\phi_2=\pi/2$ and $\phi_2=-\pi/2$, which leads to $c_0=1$, and $c_1(\alpha=0)=c_1(\alpha=\pi/2)=1$. The best numerical values of the phases that we have obtained for the cases $N=5$, $7$ and 9 are given in Tab.~\ref{Table3}. The efficiency of the composite sequences is displayed in Fig.~\ref{fig5}. Note that, for $N=5$, the pulse is not very robust along the lines $\alpha=0$ and $\alpha=\pi/2$. This is mainly due to the fact that only approximate solutions of Eq.~\eqref{eqjp2} have been derived. However, we observe that the robustness is gradually enhanced along both lines $\alpha=0$ and $\alpha=\pi/2$ as $N$ increases. Finally, we point out that the same procedure could be applied to obtain a robust pulse along more than two directions in the $(\delta,\eta)$- space, or equivalently along more than two values of the parameter $\alpha$.

\begin{table}[htp]
\caption{Same as Tab.~\ref{Table1} but for a robustness along both lines $\alpha=\pi/2$ and $\alpha=0$.}
\begin{center}
\begin{tabular}{|l|c|}
 \hline
  \hline
  $N $& $\phi_k, k=1...N$ \quad (rad) \\
   \hline
  3 & $\pi/2,  \pi/2, -\pi/2$  \\

     5 &$0.616, -0.634, 2.34, 2.58, 0.616$  \\

       7 &$1.809, 0.997, -0.292, 2.609, 2.85, 0.997, -1.333 $ \\

  9&$2.588, -0.428, 0.207, 1.803, 1.571, 4.48, 2.936, 0.428, 0.553$ \\
     \hline
      \hline
       \end{tabular}
\end{center}
\label{Table3}
\end{table}

\subsection{Robustness along all directions}\label{sec3d}
\begin{figure}[!h]
  \includegraphics[scale=0.4]{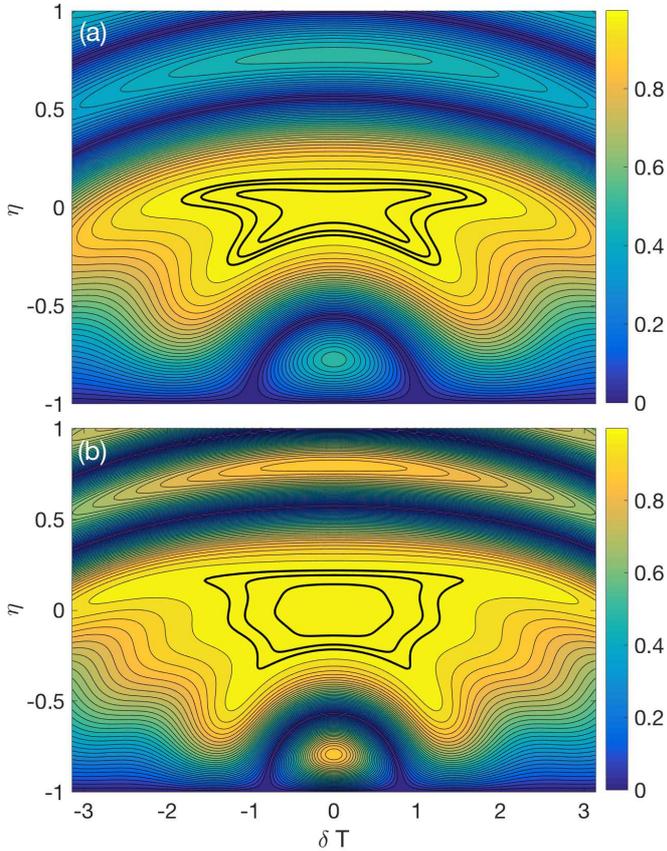}
   \caption{(color online) Same as Fig.~\ref{fig3} but for robust pulses along all directions in the space $(\delta ,\eta)$. The panels (a) and (b) correspond respectively to $N=5$ and $N=7$. Dimensionless units are used.}
  \label{fig6}
  \end{figure}

The goal of this paragraph is to generalize the preceding studies by considering a control process robust against all directions in the $(\delta,\eta)$- space. In other words, the phases of the CP must be chosen so that the values of the coefficients of the polynomial given by Eq.~\eqref{coeff} are verified for any value of $\alpha$. A solution to this problem can be derived if the coefficients $c_k$ can be expressed in the form:
\begin{equation}\label{eqtrigo}
c_k(\alpha,\phi_n)=f(\alpha)g(\phi_n)+h(\phi_n),
\end{equation}
where $f(\alpha)$ is a trigonometric function that depends only on $\alpha$ and $g(\phi_n)$ and $h(\phi_n)$ are two functions of the phases $\phi_n$. For a given value $C$, the relation $c_k(\alpha,\phi_n)=C$ is satisfied for any value of $\alpha$ if $g(\phi_n)=0$ and $h(\phi_n)=C$. As an illustrative example, we consider the case $N=5$. We consider symmetric pulses which are sufficient to derive robust solutions. From Eq.~\eqref{eqnewton} and ~\eqref{coeff}, the figure of merit can be expressed as:
\begin{equation}
J(\alpha,p,q)=c_0p^5+c_1q^2p^3+c_2q^4p,
\end{equation}
with $c_0=1$, $c_1=5/2$ and $c_2=15/8$. We also have:
\begin{equation}
\begin{cases}
c_{0}=\sin(2\phi_{1}-2\phi_{2}+\phi_{3})\\
c_{1}=\cos(2\phi_{1})-2\sin(\phi_{1})-\sin(2\phi_{1}-\phi_{3})\\-2\cos(2\alpha) \cos(\phi_{2}-\phi_{3})\left[2\sin(\phi_{1})+1\right]\\
 c_{2}=-\sin(\phi_{3})-2\cos(2\alpha)\sin(\phi_2)-2\cos(4\alpha)\sin(\phi_1).
\end{cases}
\end{equation}
We observe that the coefficient $c_1$ can be expressed under the form given in Eq.~\eqref{eqtrigo}, but it is not the case for $c_2$. We therefore only consider the coefficient $c_1$. The phases have to satisfy the following trigonometric system:
\begin{equation}
\begin{cases}
2\phi_{1}-2\phi_{2}+\phi_{3}=\pi/2\\
 \cos(\phi_{2}-\phi_{3})\left[2\sin(\phi_{1})+1\right]=0\\
 \cos(2\phi_{1})-2\sin(\phi_{1})-\sin(2\phi_{1}-\phi_{3})=5/2.
\end{cases}
\end{equation}
A solution of this system is $\phi_{2}=2\phi_{1}$, $\phi_{3}=\pi/2+\phi_{2}$ and $\phi_{1}=-\pi/6$. Due to the complexity of the equations we have limited the computation to $N = 7$ and to symmetric pulses. For $N=7$, we have found $\phi_{1}=\phi_{4}/2-\pi/4=\phi_{7}$, $\phi_{2}=3\phi_{1}=\phi_{6}$,  $\phi_{3}=3\phi_{1}=\phi_{5}$ and $\phi_{4}=\arcsin(\sqrt{3}-19/16)$. Note that with these values, only the relations associated with the coefficients $c_1$ and $c_2$ are satisfied for $N=7$. We conjecture that the same approach can be applied to higher degree $N$, and in particular that the coefficients $c_1$, $c_2$, $\cdots$, and $c_{\frac{N-3}{2}}$ can be written in the form \eqref{eqtrigo}. The efficiency of the derived CPs is represented in Fig.~\ref{fig6}. Finally, we point out that the $5$- pulse sequence is the same as the one derived by Knill and described in~\cite{jones:2013,CPknill}.
  \subsection{Comparison with other composite pulses}\label{sec3e}
\begin{figure}[!h]
  \includegraphics[scale=0.32]{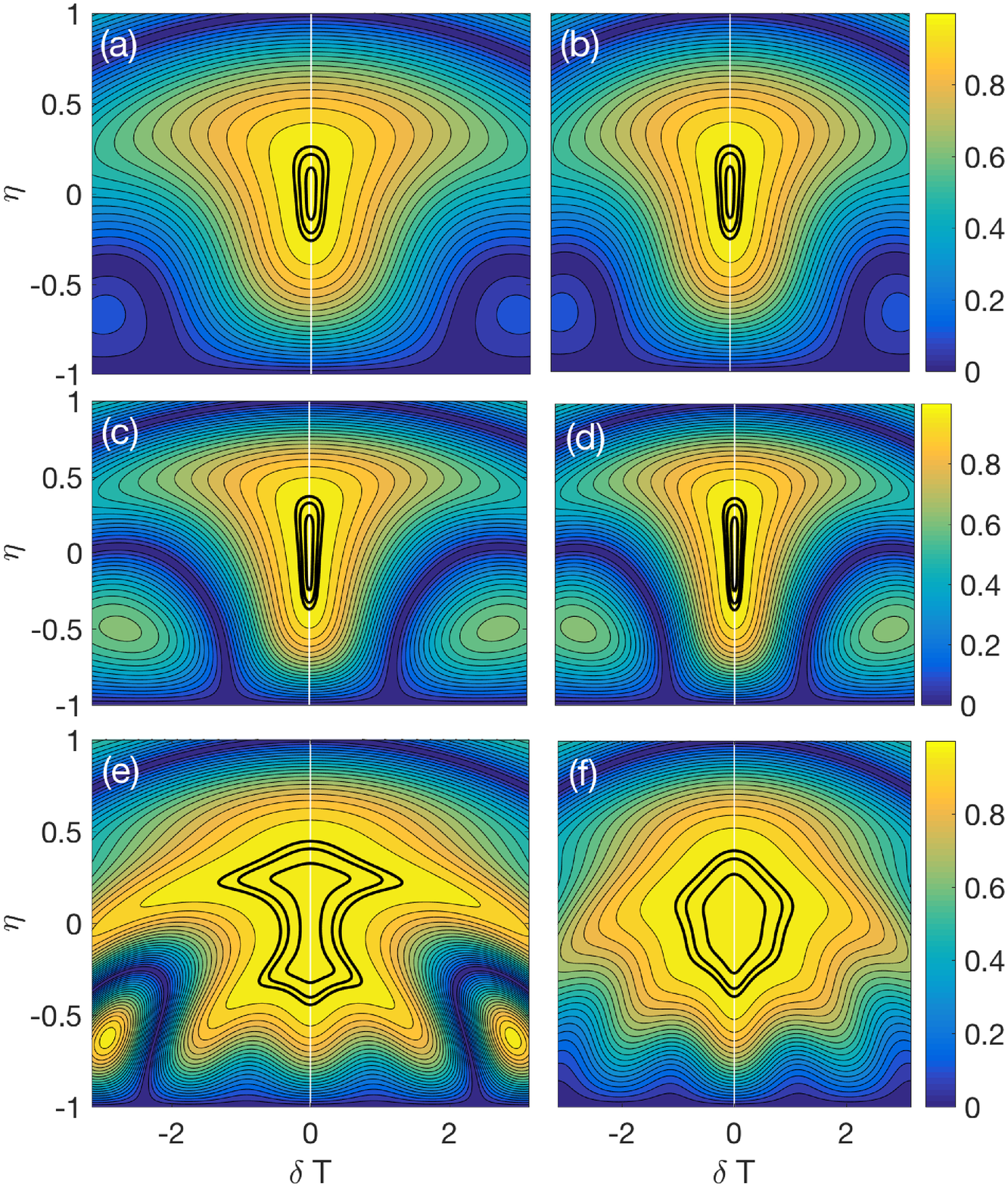}
   \caption{(color online) Comparison between the CPs derived in this study (panels (a), (c) and (e)) and the ones of Ref.~\cite{jones:2013} (panels (b), (d) and (f)). We consider pulses robust along the $\alpha=0$- direction. The first, second and third rows correspond respectively to the orders $N=3$, $N=5$ and $N=7$.  The black bold solid lines (from inside to outside) correspond respectively to the level lines $\vert J(\delta ,\eta) \vert$=0.999, 0.995 and 0.99. Dimensionless units are used.}
  \label{comparison1}
  \end{figure}

\begin{figure}[!h]
  \includegraphics[scale=0.275]{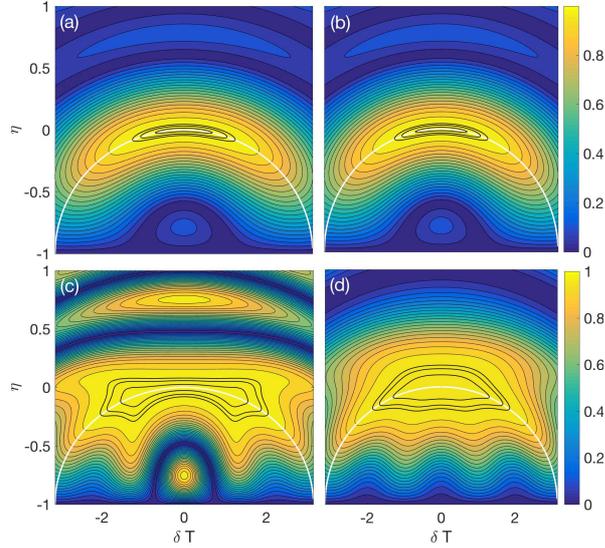}
   \caption{(color online) Same as Fig.~\ref{comparison1} but for robust pulses along the $\alpha=\pi/2$- direction. Dimensionless units are used.}
  \label{comparison2}
  \end{figure}

\begin{figure}[!h]
     \centering
  \includegraphics[scale=0.3]{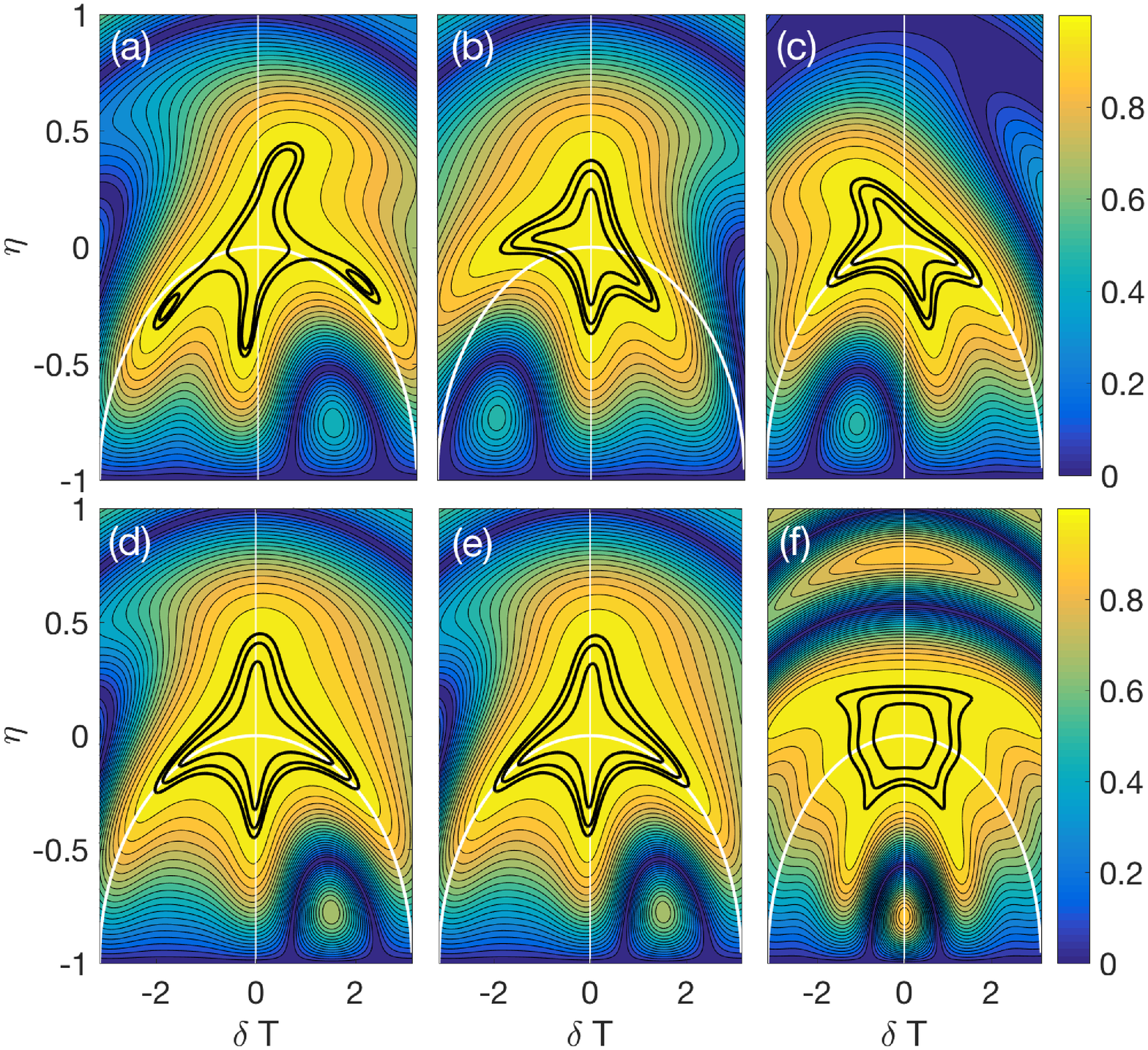}
   \caption{(color online)  Same as Fig.~\ref{comparison1} but for robust pulses along both directions.  The first and second rows correspond respectively to $N=5$ and $N=7$. Panel (a) displays the performance of a CP with $N=5$ derived in this study. Panels (b) and (c) correspond to CPs of Ref.~\cite{jones:2013} with $N=5$ and designed to suppress respectively the second order pulse strength and second order off-resonance errors. Panels (d) and (e) represent the effect of a CP of Ref.~\cite{jones:2013} with $N=7$. We add for comparison in the panel (f) a figure corresponding to a 7-$\pi$ CP designed in this study to be robust along all directions. Dimensionless units are used.}
  \label{comparison3}
  \end{figure}

In this section, we compare the CPs derived in this study with the ones obtained in Ref.~\cite{jones:2013} by using a different and more geometric approach. Note that a NOT gate of the form
   \begin{equation}
   U_{t}=\left( \begin{matrix}
  0 & -i\\
  -i & 0
  \end{matrix}\right).
  \end{equation}
is considered in~\cite{jones:2013}, i.e. a rotation of $\pi$ around the $x$- axis. In this case, the pulse phases should fulfill $\sum_{j=1}^{N}(-1)^{j+1}\phi_{j}=0$. For the NOT gate of Eq.~\eqref{Eq.11} which corresponds to a $\pi$- rotation around the $y$- axis, they satisfy $\sum_{j=1}^{N}(-1)^{j+1}\phi_{j}=\pi/2$.
Figures~\ref{comparison1},~\ref{comparison2} and~\ref{comparison3} show a systematic comparison between the different CPs for a robustness respectively along the $\alpha=0$, $\alpha=\pi/2$ or along both directions. We recall that $\alpha=0$ corresponds to pulse strength error in~\cite{jones:2013}, while $\alpha=\pi/2$ is associated to off-resonance errors. We observe that the pulses for $N=3$ and $N=5$ are the same as~\cite{jones:2013}, even if they have been derived with a distinct approach. The CPs for $N=7$ are different and we note that our CP is more robust along the line $\delta=0$. The same observation can be made in Fig.~\ref{comparison2} at the order $N=7$. Here again, we stress the slightly better robustness of our CP along the $\alpha=\pi/2$- direction. For the robustness along two directions, only the CP at the order 7 is the same as the one derived in~\cite{jones:2013}. The other solutions are different even if the corresponding efficiency is very similar for the same number of individual pulses. The performance of CPs along both directions is shown in Fig. ~\ref{comparison3}.


\section{Conclusion and prospective views}\label{conc}
In this paper, we have presented a general procedure for the design of CPs robust against both offset uncertainties and control field variations. The NOT gate is taken as an illustrative example. We have shown that the robustness can be achieved along specific lines of this two-dimensional space or along all directions. In this approach, the construction of robust control strategies is replaced by the determination of solutions of a nonlinear trigonometric system. The robustness can be achieved with a pulse depending upon a very small number of parameters. This is in contrast with the large number of parameters (typically a few thousand) needed in optimal control techniques~\cite{kobzar2004}. In addition, the determination of the solutions of a nonlinear system is faster in terms of computational time than optimization processes. For a robustness along a single direction defined by a constant value of $\alpha$, we have shown that a symmetric composite sequence is solution of the nonlinear system. However, only numerical solutions can be derived if two or more lines are accounted for. We have also generalized this approach to all directions in the ($\delta , \eta$)- space. Note that, in this latter situation, a symmetric CP is used. In this case, the figure of merit has $(N+1)/2$ coefficients with $N$ phases $\phi_{n}$ to determine. To handle this non-square system, we can use for instance the Levenberg-Marquardt algorithm~\cite{Gill}. The control strategy developed in this paper could be generalized to any gate in SU(2). The derivation of efficient pulses requires however a more involved parametrization of the pulses in which both the amplitude and the phase can be adjusted. This point which goes beyond the scope of this study will be the subject of a forthcoming paper.

Recent experimental studies in quantum computing~\cite{genov:2014,vandamme:2017} and in NMR~\cite{skinner2003,Nimbalkar2013,kobzar2005} show that the pulses derived in this paper could be implemented experimentally. Complex pulse sequences can by now be used with modern NMR spectrometers. More precisely, the amplitude  and phase of the control pulses can be defined with high resolution, allowing for a virtually continuous variation of these parameters~\cite{NMRexp1,NMRexp2}. A good match between theory and experiment is generally observed.

\noindent\textbf{ACKNOWLEDGMENT}\\
D. Sugny acknowledges support from the PICS program, from the ANR-DFG research program COQS (ANR-15-CE30-0023-01) and from the QUACO project (ANR 17-CE40-0007-01). This project has received funding from the European Union's Horizon 2020 research and innovation programme under the Marie-Sklodowska-Curie grant agreement No 765267 (QUSCO).

\appendix
\section{General expression of the coefficients $A_k, B_k, C_k,$ and $D_k$ for $N=5$}\label{coeffapp}
In this Appendix, we give  the general expression of the coefficients $A_k, B_k, C_k,$ and $D_k$ for $N=5$. Using Eq.~\eqref{eq_pq} and after straightforward simplifications, we obtain:
$\Re{\left( U^{(N)}_{1,1}\right)}=\sum_{k=0}^{2} A_{k}p^{2k}q^{N-2k}$ with,
\begin{eqnarray*}
A _0 &=&\cos(5\alpha),  \\
A_1&=&-\cos(\alpha-\varphi_2+\varphi_5)-\cos(-\varphi_1+\alpha+\varphi_4)\\
&-&\cos(\varphi_1+\alpha-\varphi_3) -\cos(-\varphi_1+3\alpha+\varphi_5)\\
&-&\cos(3\alpha+\varphi_2-\varphi_3)-\cos(3\alpha-\varphi_5+\varphi_4)\\
&-&\cos(3\alpha+\varphi_3-\varphi_4)-\cos(\varphi_1-\varphi_2+3\alpha)\\
&-&\cos(\alpha+\varphi_2-\varphi_4)-\cos(\alpha+\varphi_3-\varphi_5), \\
A_2&=&\cos(-\varphi_1+\alpha+\varphi_3+\varphi_5-\varphi_4)\\
&+&\cos(-\varphi_1+\varphi_2-\varphi_3+\varphi_5+\alpha)\\
&+&\cos(\varphi_1-\varphi_2+\varphi_3+\alpha-\varphi_4)\\
&+&\cos(\alpha+\varphi_2-\varphi_3-\varphi_5+\varphi_4)\\
&+&\cos(\varphi_1-\varphi_2+\alpha-\varphi_5+\varphi_4),
\end{eqnarray*}
$\Im{\left( U^{(N)}_{1,1}\right)}=\sum_{k=0}^{2} B_{k}p^{2k}q^{N-2k}$ with:
\begin{eqnarray*}
B_0 &=&\sin(5\alpha), \\
B_1&=&-\sin(\varphi_1+\alpha-\varphi_3)-\sin(3\alpha+\varphi_3-\varphi_4) \\
&-&\sin(\alpha+\varphi_3-\varphi_5)+\sin(-\varphi_1+3\alpha+\varphi_5) \\
&-&\sin(\alpha+\varphi_2-\varphi_4)+\sin(\alpha-\varphi_2+\varphi_5)\\
&-&\sin(\varphi_1-\varphi_2+3\alpha)+\sin(-\varphi_1+\alpha+\varphi_4)\\
&-&\sin(3\alpha-\varphi_5+\varphi_4)-\sin(3\alpha+\varphi_2-\varphi_3), \\
B_2&=&-\sin(-\varphi_1+\varphi_2-\varphi_3+\varphi_5+\alpha)\\
&+&\sin(\alpha+\varphi_2-\varphi_3-\varphi_5+\varphi_4)\\
&+&\sin(\varphi_1-\varphi_2+\varphi_3+\alpha-\varphi_4)\\
&+&\sin(\varphi_1-\varphi_2+\alpha-\varphi_5+\varphi_4)\\
&-&\sin(-\varphi_1+\alpha+\varphi_3+\varphi_5-\varphi_4),
\end{eqnarray*}
$\Re{\left( U^{(N)}_{2,1}\right)}=\sum_{k=0}^{2} C_{k}q^{2k}p^{N-2k}$ with:
\begin{eqnarray*}
C_0 &=&\sin(\varphi_1-\varphi_2+\varphi_3+\varphi_5-\varphi_4), \\
C_1&=&-\sin(\varphi_1-\varphi_2+\varphi_4)+\sin(-\varphi_1+\varphi_2-\varphi_3+2\alpha)\\
&-&\sin(\varphi_1-\varphi_3+\varphi_5)+\sin(-\varphi_1+2\alpha-\varphi_5+\varphi_4)\\
&-&\sin(\varphi_2-\varphi_3+\varphi_4)-\sin(\varphi_2+\varphi_5-\varphi_4)\\
&+&\sin(-\varphi_1+2\alpha+\varphi_3-\varphi_4)-\sin(2\alpha+\varphi_2-\varphi_3+\varphi_5)\\
&-&\sin(2\alpha+\varphi_3+\varphi_5-\varphi_4)-\sin(\varphi_1-\varphi_2+2\alpha+\varphi_5), \\
C_2&=&\sin(2\alpha+\varphi_4)-\sin(-\varphi_1+4\alpha) \\
&+&\sin(4\alpha+\varphi_5)+\sin(\varphi_3)-\sin(2\alpha-\varphi_2),
\end{eqnarray*}
and $\Im{\left( U^{(N)}_{2,1}\right)}=\sum_{k=0}^{2} D_{k}q^{2k}p^{N-2k}$, with:
\begin{eqnarray*}
D_0 &=&\cos(\varphi_1-\varphi_2+\varphi_3+\varphi_5-\varphi_4), \\
D_1&=&-\cos(\varphi_2-\varphi_3+\varphi_4)-\cos(-\varphi_1+\varphi_2-\varphi_3+2\alpha) \\
&-&\cos(\varphi_1-\varphi_2+\varphi_4)-\cos(-\varphi_1+2\alpha-\varphi_5+\varphi_4) \\
&-&\cos(\varphi_2+\varphi_5-\varphi_4)-\cos(\varphi_1-\varphi_3+\varphi_5) \\
&-&\cos(-\varphi_1+2\alpha+\varphi_3-\varphi_4)-\cos(\varphi_1-\varphi_2+2\alpha+\varphi_5),\\
&-&\cos(2\alpha+\varphi_2-\varphi_3+\varphi_5)-\cos(2\alpha+\varphi_3+\varphi_5-\varphi_4)\\
D_2&=&\cos(-\varphi_1+4\alpha)+\cos(\varphi_3)\\
&+&\cos(4\alpha+\varphi_5)+\cos(2\alpha-\varphi_2)+\cos(2\alpha+\varphi_4).
\end{eqnarray*}
\bibliographystyle{apsrev}.

\end{document}